\documentclass[letter,onecolumn]{jpsj3}
\usepackage{graphicx}
\usepackage{latexsym}
\usepackage{fancybox}
\usepackage{color}

\title{Superconductivity in the Ferroquadrupolar State in the Quadrupolar Kondo Lattice PrTi$_2$Al$_{20}$ }

\author{Akito Sakai, Kentaro Kuga and Satoru Nakatsuji\thanks{E-mail: satoru@issp.u-tokyo.ac.jp}}  
\inst{Institute for Solid State Physics, University of Tokyo, Kashiwa, Chiba 277-8581, Japan}

\recdate{\today}

\abst{ The cubic compound PrTi$_2$Al$_{20}$  is a quadrupolar Kondo lattice system that exhibits quadrupolar ordering due to the non-Kramers $\Gamma_3$ ground doublet and has strong hybridization between $4f$ and conduction electrons. Our study using high-purity single crystal reveals that  PrTi$_2$Al$_{20}$ exhibits type-II superconductivity at $T_{\rm c} = 200$ mK in the nonmagnetic ferroquadrupolar state. The  superconducting critical temperature and field phase diagram indicates clean limit superconductivity with moderately enhanced effective mass of $m^*/m_0 \sim 16$. 
}

\kword{Pr$Tr_2$Al$_{20}$, quadrupolar Kondo lattice, nonmagnetic Kondo effect, quadrupolar order}
\begin{document}
\maketitle
\newpage
Orbital degree of freedom in strongly correlated electron systems leads to rich variety of nontrivial phenomena, such as colossal magnetoresistance, orbital ordering \cite{tokuranagaosa}, and spin-orbital disordered states \cite{BCSO}. In  particular, for the study of topical subjects of $d$-electron systems including high-temperature superconductivity and frustrated magnetism, orbital arrangement has been found to provide an important basis to understand the underlying electronic and spin structures. This is because of the strong coupling between orbital, spin and charge degrees of freedom, which, on the other hand, makes it hard to study purely orbital-driven effects. 

Interestingly, however, $f$-electron systems may provide a rare example of {\it nonmagnetic} ground states that have mainly orbital degree of freedom known as quadrupolar moments, i.e. irreducible tensor operators of the total angular momentum {\boldmath{$J$}} \cite{Kuramoto}. In particular,  cubic Pr and U based compounds with $4f^{2}$  configuration are known to sometimes exhibit a nonmagnetic ground state with the $\Gamma_3$ quadrupolar degree of freedom.   In comparison with the uranium compounds, Pr based materials have been found to exhibit rich variety of electronic ground state due to quadrupole moments, such as ferro- and antiferroquadrupolar states, e.g. in PrPtBi and  PrPb$_3$ \cite{RefWorks:364,Morin1982257,Onimaru}, and superconductivity in a quadrupolar phase in PrIr$_2$Zn$_{20}$  \cite{Onimaru2,Onimaru3}.

Moreover, for these nonmagnetic states, the hybridization between the $4f$ and conduction electrons is expected to lead to two competing interactions. One is the indirect RKKY type intersite coupling between quadrupole moments, which may sometimes induce incommensurate modulated structure as observed in PrPb$_3$ \cite{Onimaru}. The other is the two-channel (quadrupolar) Kondo effect that quenches the orbital degrees of freedom and stabilize an anomalous metallic state \cite{RefWorks:366,CoxPhysica,Kusunose}.
For Ce and Yb based compounds described as Kondo lattice systems, it is known that the competition between the Kondo effect and RKKY interaction gives rise to quantum criticality and related phenomena such as unconventional superconductivity and non-Fermi liquid behavior. In analogy, novel quantum criticality and exotic superconductivity might arise as a result of the competition between the quadrupolar Kondo effect and RKKY interaction.

In the search for the novel screening effects, pioneer works on the cubic $4f^{2} \Gamma_3$ systems PrInAg$_2$ and PrMg$_3$ have found that these compounds do not exhibit any long-range order \cite{RefWorks:71,RefWorks:335} and suggested a nonmagnetic heavy fermion state due to the quadrupolar Kondo effect. However, both are Heusler-type compounds where a random site exchange may exist and lift the degeneracy of the non-Kramers $\Gamma_3$ doublet that is essential for the nonmagnetic Kondo effect. 

On the other hand, existence of the quadrupolar ordering guarantees the orbital degeneracy above its ordering temperature. Therefore, it is ideal if we may decrease the ordering temperature by increasing the hybridization strength, because it may allow us to study the competition effects between nonmagnetic Kondo effects and the quadrupolar order, keeping the degeneracy of quadrupolar degree of freedom. Recently, we have found that a series of cubic $4f^{2} \Gamma_3$ compounds, Pr$Tr_2$Al$_{20}$ ($Tr$: transition metal),  are such {\it quadrupolar Kondo lattice systems} that allow us to tune the hybridization strength and the quadrupolar ordering temperature \cite{akito}. Among them, PrTi$_2$Al$_{20}$ is the best studied and established to have ferroquadrupolar order as well as strong hybridization between 4$f$ and conduction electrons.

The crystal structure of PrTi$_2$Al$_{20}$ is cubic CeCr$_2$Al$_{20}$-type with the space group $Fd{\bar 3}m$ \cite{RefWorks:358}. In particular, the symmetry of $R$ site is $T_{\rm d}$ and cubic. High coordination number of 16 Al ions around Pr leads to strong hybridization between $4f$ and conduction electrons \cite{RefWorks:360}. Indeed, as a rare case in Pr based compound, the Kondo effect has been observed such as  the $-\ln T$ dependence of the $4f$ component of the resistivity  \cite{akito}. Furthermore, resonant photoemission spectroscopy has revealed the Kondo resonance at the Fermi level \cite{Matsunami}.  On the other hand, our thermomagnetic measurements found that the Pr ions have the nonmagnetic $\Gamma_3$ doublet and exhibit quadrupolar ordering  at  $T_{\rm Q} = 2.0 $ K.  Inelastic neutron experiments confirmed that the crystal electric field (CEF) scheme has the ground doublet of $\Gamma_3$, which is separated from the excited levels, $\Gamma_4$ by 5.7 meV, $\Gamma_5$ by 9.5 meV, and $\Gamma_1$ by 13.6 meV \cite{Sato}. It also clarified that the quadrupolar phase has a ferro-type uniform order of  the quadrupolar moment $O^{0}_2$, consistent with  the ferro-type intersite coupling found in the ultrasonic measurements \cite{Nakanishi}, and the absence of internal magnetic field confirmed by recent $\mu$SR measurements \cite{Ito}.  Hybridization effects between the quadrupolar moments and conduction electrons are inferred  from  the magnetic properties  above the ordering temperatures, such as the $-T^{1/2}$ dependence of the susceptibility and $T^2$ dependence of the resistivity  \cite{akito}. 

Here, we report the observation of the superconductivity at $T_{\rm c} = 200 $~mK in the ferroquadrupolar state of PrTi$_2$Al$_{20}$.  The field dependence of the critical temperature indicates type-II superconductivity and the moderately enhanced effective mass of $m^*/m_0 \sim 16$. 
This provides an interesting counterpart to the superconductivity at $T_{\rm c} = 50 $~mK found in the antiferro-quadrupolar state in the isostructural compound PrIr$_2$Zn$_{20}$  \cite{Onimaru2,Onimaru3}, which in contrast has well localized $4f$ electrons.
More enhanced critical temperature and field in PrTi$_2$Al$_{20}$  should come from the mass enhancement due to the strong hybridization between $4f$ and conduction electrons.

\begin{figure}[t]
\begin{center}
\includegraphics[keepaspectratio, scale=1.4]{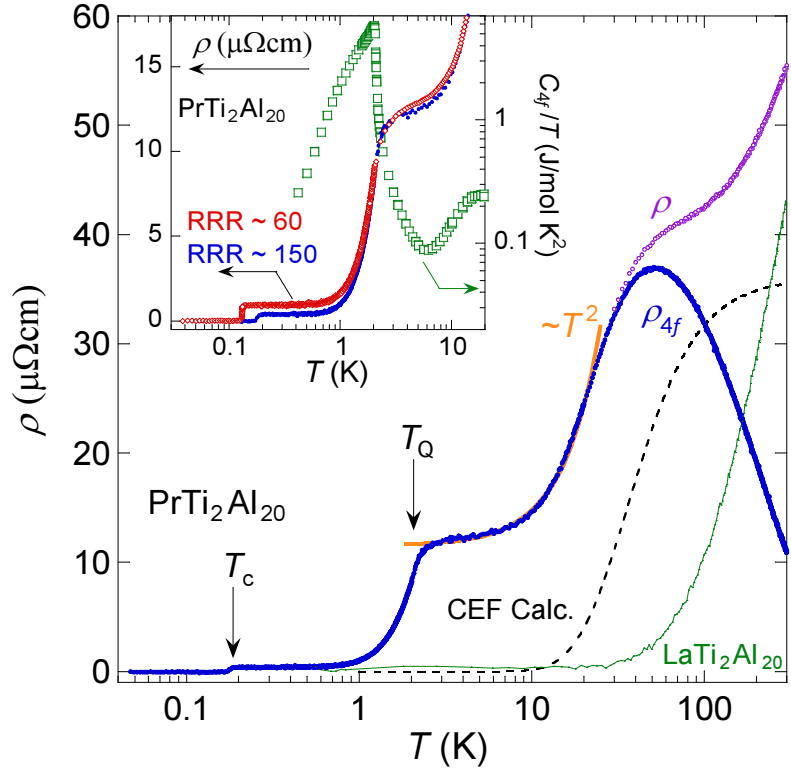}
\caption{(Color online) Temperature dependence of the resistivity  $\rho(T)$ (open circle) and $4f$ electron contribution to the resistivity $\rho_{4f}(T)$ (solid circle) of PrTi$_2$Al$_{20}$ and $\rho(T)$ of LaTi$_2$Al$_{20}$ (green solid line) under the earth field. The error in estimating $\rho$ is less than 10 $\%$. Arrows indicate the ferro-quadrupole transition temperature $T_{\rm Q} = 2.0 $ K and superconducting transition temperature $T_{\rm c}$ = 0.2 K, respectively. Orange solid line represents the $T^2$ fitting. Black broken line indicates $\rho(T)$ calculated using the CEF parameters (arbitrary unit) determined by inelastic neutron scattering experiment \cite{Sato}. Inset left axis: $\rho(T)$ below 20 K for two single crystals with RRR of $\sim$150 (solid circle) and $\sim$ 60 (open diamond). Inset right axis: $4f$ contribution to the specific heat divided by temperature, $C_{4f}/T$ (open square).
}\label{Rho}
\end{center}
\end{figure}

Single crystals of $R$Ti$_2$Al$_{20}$ ($R=$Pr, La) were grown by an Al self-flux method in a sealed quartz tube, using 4N(99.99\%)-Pr, 3N-La, 3N-Ti and 5N-Al \cite{akito}. Single and powder X-ray diffraction measurements found the single phase. For the resistivity measurements, long thin plates are prepared to reduce the error $(< 10 \%)$ in calculating the resistivity and an ac four-probe method has been employed. The average value at room $T$ resistivity $\rho(300$ K$)$  is found consistent with the one reported previously \cite{akito} and is used to normalize $\rho(300$ K). The ac-susceptibility was measured by a mutual inductance method with an ac field (0.6 mOe, 1.2 kHz) using a dilution refrigerator at 0.03 K $\leq T \leq $ 2 K. 
We also utilized the ac and dc signal output of a SQUID probe in the dilution refrigerator measured under a respective ac field (1 mOe, 32 Hz) and dc field (100 mOe)  applied along [110]. For the SQUID measurements, the earth's magnetic field was eliminated by using a Nb superconducting (SC) shield covered with a $\mu$-metal tube.
As a reference superconductor, Al was put in the canceling coil wound in reverse to the pickup coil used for PrTi$_2$Al$_{20}$.
The specific heat $C_P$ down to $T = 0.4$ K was measured by a thermal relaxation method. The $4f$ contribution, $C_{ 4f}$, was estimated by subtracting $C_P$ of the La analog, LaTi$_2$Al$_{20}$ from $C_P$ of PrTi$_2$Al$_{20}$.

Figure \ref{Rho} shows the $T$ dependence of the resistivity, $\rho(T)$, at zero field (open symbol), and its $4f$ electron contribution $\rho_{4f}(T)$ (solid symbol), obtained after subtracting $\rho(T)$  of the La analog (solid line). 
On cooling from room $T$, $\rho_{4f}$ clearly exhibits a $-\ln T$ behavior and forms a peak at $\sim 60$ K, corresponding to the CEF gap $\sim 60$ K between the ground $\Gamma_3$ and the excited $\Gamma_4$ states. 
This $-\ln T$ behavior always appears within the error bar of the resistivity.
$\rho (T)$ calculated in Born approximation \cite{SpinDisorderResV} by using the CEF parameters determined by inelastic neutron scattering experiment\cite{Sato} (broken line, arbitrary unit) only saturates to a constant value at high $T$ and thus the $-\ln T$ dependence should come from the Kondo effect \cite{akito}.
On further cooling, $\rho_{4f}$ shows $T^2$ dependence below 20 K where $\Gamma_3$ quadrupolar degree of freedom dominates (solid line in Fig. \ref{Rho}).  In the same $T$ range, the $4f$ electron contribution to the specific heat divided by temperature, $C_{4f}/T$, shows nearly constant behavior with the value $\gamma \sim 100$ mJ/mole K$^2$ (Inset of Fig \ref{Rho}, open square and right axis).  Using the $T^2$ coefficient of $A \sim 3.3$ $\times 10^{-2}$ $\mu\Omega$cm/K$^2$ of $\rho_{4f}(T)$ ,  the Kadowaki-Woods ratio $A/\gamma^2$ is estimated to be $\sim 3.3 \times 10^{-6} \mu\Omega$cmK$^2$mole$^2$/mJ$^2$, which is close to the universal value of $1 \times 10^{-5}\mu\Omega$cmK$^2$mole$^2$/mJ$^2$ known for strongly correlated electron systems \cite{Kadowaki-Woods}. 
This suggests that the $\gamma$ value comes from the mass enhancement due to the hybridization between the quadrupolar moments and conduction electrons.

\begin{figure}[t]
\begin{center}
\includegraphics[keepaspectratio, scale=2.8]{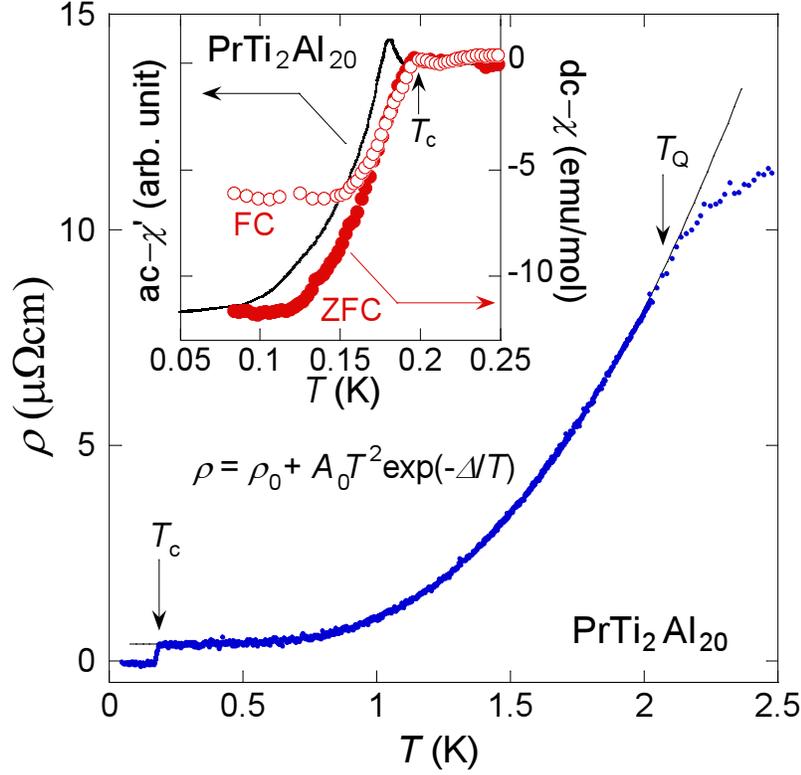}
\caption{(Color online) Temperature dependence of the resistivity $\rho(T)$ below 2.5 K for the same crystal with RRR = 150 in Fig. \ref{Rho} (solid circle).  Arrows indicate the ferro-quadrupole transition temperature $T_{\rm Q} = 2.0 $ K and superconducting transition temperature $T_{\rm c}$ = 0.2 K. Solid line represents the fitting to the formula, $\rho(T)= \rho_0+A_0T^2\exp (-\Delta /T)$. Inset left axis: $T$ dependence of the real part of the ac-susceptibility $\chi^{\prime}$ (solid line) measured by a mutual-inductance method under ac-field of 0.60 mOe with frequency of 1.2 kHz. 
Differential paramagnetic effect is observed at $\sim T_{\rm c}$ due to the earth field.
Inset right axis: $T$ dependence of the dc-susceptibility $\chi$ for zero field cooled (ZFC, solid circle) and field cooled (FC, open circle) under a field of 0.1 Oe.
}\label{RhoLow}
\end{center}
\end{figure}

A clear drop is observed in $\rho(T)$  of PrTi$_2$Al$_{20}$ owing to the ferroquadrupole order at  $T_{\rm Q} = 2.0 $ K. Strikingly, with further decreasing $T$ in this ferroquadrupolar phase, PrTi$_2$Al$_{20}$ is found to exhibit superconductivity; the resistivity vanishes to zero at $T_{\rm c}\sim 0.2$ K. The particular single crystal used for the main frame of  Fig. \ref{Rho} has the residual resistivity ratio (RRR) of  $\sim 150$ and residual resistivity of $\rho_0 = 0.38 $ $\mu\Omega$m, indicating ultrapurity of the sample. The lower quality sample with RRR = 60 has a lower $T_{\rm c}$ = 140 mK. Above 1.5 K, on the other hand, $\rho(T)$ for both samples collapses on top, indicating that the quadrupolar transition temperature,  $T_{\rm Q}$, does not change with RRR (Inset of Fig. \ref{Rho} left axis).
In the ferroquadrupolar phase below $T_{\rm Q}$, $\rho(T)$ shows an exponential decrease over a decade of $T$ down to $T_{\rm c}\sim 0.2$ K (left axis and blue solid circle of Fig.  \ref{RhoLow}), and can be well fit by $\rho = \rho_0 + A_0T^{2}\exp(-\Delta/T)$ with $\rho_0 = 0.38\  \mu\Omega$cm, $A_0 = 5.6$ $\mu\Omega$cm$/K^2$ and $\Delta = 2.1$ K (solid curve in Fig. \ref{RhoLow}). This suggests the emergence of the anisotropy gap of the collective mode due to the uniform order of the quadrupolar moment $O^{0}_2$ below $T_{\rm Q}$, suppressing the orbital fluctuations.
 Note that $\rho(T)$ of lower quality sample can also be well fit with almost the same parameters except $\rho_0$ ($\rho_0=0.93$ $\mu\Omega$cm, $A_0 \sim 5.1 \mu\Omega$cm/K$^2$,  $\Delta = 2.0$ K), consistent with the robustness of the quadrupole order.


To confirm the bulk nature of the superconductivity, we have performed both ac- and dc- susceptibility measurements.
Inset of Fig. \ref{RhoLow} shows the $T$ dependence of the real part of the ac-susceptibility $\chi'$ (solid line, left axis) and dc-susceptibility $\chi$ (right axis) taken under zero field cooled (ZFC) and field cooled (FC) sequences. 
Measurements were made under the earth field for ac-$\chi'$ and under a field of 0.1 Oe for dc-$\chi$ (using the SC shield).
The dc-$\chi$ data is corrected by using demagnetization factor $N \approx 0.2$ estimated for the rectangular bar shape of the crystal \cite{demag}. 
A clear diamagnetic signal due to the SC transition was found below the onset temperature of $T_{\rm c}= 0.2$ K. 
For dc-$\chi(T)$, the FC (open circle) and ZFC (solid circle) results clearly bifurcate below $T_{\rm c}$, while the ac-$\chi'$ show no hysteresis.
The SC volume fraction is estimated to be $\sim 60\%$ (ZFC) and $\sim 35\%$ (FC) by using the value of dc-$\chi$ at the lowest $T$ measured. 
We also confirmed that the diamagnetic signal in the ac-$\chi$ of PrTi$_2$Al$_{20}$ has the similar size to the one (not shown) of the Al reference at 1.2 K, which has nearly the same shape and volume as the PrTi$_2$Al$_{20}$ crystal. These results confirm the bulk nature of the superconductivity.

\begin{figure}[t]
\begin{center}
\includegraphics[keepaspectratio, scale=2.8]{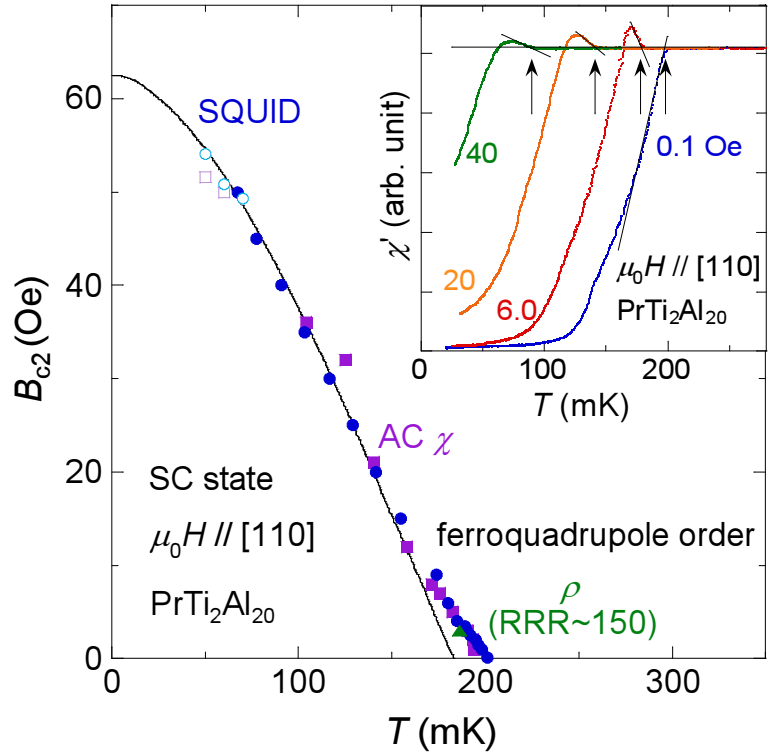}
\caption{(Color online) Temperature dependence of the upper critical field $B_{\rm c2}$ along [110] axis. The square/circle data points are determined by mutual-inductance /SQUID method and solid/open symbols represent $B_{\rm c2}$ determined by temperature/field sweep, respectively. The triangle symbol indicates $B_{\rm c2}$ determined by the temperature sweep measurement of the resistivity using the sample with RRR =150. Solid line represents the fit based on the WHH model (see text). Inset: Temperature dependence of the real part of ac-susceptibility $\chi^{\prime}$ measured using SQUID under various magnetic fields. The earth field was eliminated by shields. The critical temperatures (arrows) were determined as the onset of the differential paramagnetic effect. }\label{HT}
\end{center}
\end{figure}

Field suppression of the superconductivity was investigated using the ac-susceptibility measurements.
Figure \ref{HT} shows the $T$ dependence of the upper critical field $B_{\rm c2}$ of the SC phase. Here, the critical temperature and field were defined as the onset of the anomaly of the susceptibility as a function of temperature (solid symbol in Fig. \ref{HT}) and field (open symbol) (Inset of Fig. \ref{HT}).  
The peak around $T_{\rm c}$ observed under $B > 0.1$ Oe is attributable to differential paramagnetic effect (DPE) \cite{DPE}, which implies that pinning of magnetic flux in the mixed state is weak because of high quality of the crystal. The height of DPE is much smaller than diamagnetic signal of Meissner effect, consistent with the type-II superconductivity, which will be discussed below. The extrapolation of the  $T$ dependence of $B_{\rm c2}$ provides a rough estimate of the zero temperature values, $B_{\rm c2}(0) \sim 6$ mT. 

Two pair-breaking mechanisms are known as the origin of the critical field, namely, orbital depairing and paramagnetic effects. First, orbital critical fields can be evaluated based on the Werthamer-Helfand-Hohenberg (WHH) model  \cite{WHH,HW}. 
The model well fits the observed curve below $\sim$150 mK (Fig. \ref{HT}, solid curve). Because of the concave curve found above $T\sim$150 mK, the best fit was obtained with the following parameters: a slightly lower $T_{\rm c}$ = 0.183 K  and a higher gradient at $T_{\rm c}$,  $B_{\rm c2}^{\prime} \equiv dB_{\rm c2}/dT \sim -47$ mT/K than those observed in experiment. 
The fitting yields the orbital critical field at absolute zero in the clean limit, $B_{\rm c2}^{\rm orb}(0) = -0.727 B_{\rm c2}^{\prime}T_{\rm c} = 6.3$ mT.   On the other hand, the Pauli limit due to the paramagnetic effect is estimated by $B_{\rm c2}^{\rm Pauli}(0) = 1.83 T_{\rm c}$ \cite{Clogston}. This provides $B_{\rm c2}^{\rm Pauli}(0)= 335$ mT, which is one order magnitude larger than the observation. 

Interestingly, $B_{\rm c2}(T)$ in the $T$ range of $\sim150$ mK $<T<T_{\rm c}$ slightly deviates from the curve based on the WHH model and shows concave feature with positive $d^2B_{\rm c2}/dT^2$. Such concave curve of $B_{\rm c2}(T)$ has also been observed in various superconductors, and has been discussed theoretically based on various mechanisms. 
For the superconductivity of PrTi$_2$Al$_{20}$, which is in clean limit as we will discuss, and appears in the ferroquadrupolar state in a 3D compound, the multi-band effect is a likely origin \cite{Shulga}. 
Further measurements are necessary to confirm this multigap scenario.

Using the critical fields determined above, the Ginzburg-Landau (GL) equation, $B_{\rm c2}^{\rm orb}(0) = \Phi_0/(2 \pi \xi^2)$,  yields the coherence lengths $\xi(0) = 0.23$ $\mu$m.
We may also estimate physical parameters, following the BCS theory \cite{BCS,BCS2}. Considering the cubic structure, we assume a spherical Fermi surface with a Fermi wave number $k_{\rm F}$. Using $B_{\rm c2}^{\prime} = -47$ mT/K used for the WHH fitting, $\rho_0 = 0.38$ $\mu\Omega$ cm (RRR = 150), $T_{\rm c}=0.183$ K, and the specific heat coefficient $\gamma \sim 100$ mJ/mol K$^2$, we obtain $k_{\rm F} = 5.0 \times 10^9$ 1/m, the mean free path $l = 1.3$ $\mu$m, the coherence length $\xi_0 = 0.28$ $\mu$m, and the penetration depth $\lambda = 0.33$ $\mu$m. Thus, $\kappa = \lambda/\xi = 1.2 $ indicates type-II superconductivity. Note $\xi_0$ is roughly consistent with the values obtained by the GL equation above. Furthermore, the mean free path $l$ is one order of magnitude longer than $\xi_0$, indicating that the superconductivity is within the clean limit. 
The effective mass $m^*/m_0$ is estimated to be $\sim 16$, and this moderately enhanced value suggests the hybridization effects. Furthermore, this is consistent with the moderate enhancement of the specific heat coefficient $\gamma \sim 100$ mJ/mol K$^2$ for PrTi$_2$Al$_{20}$, in comparison with $\gamma \sim 23$ mJ/mol K$^2$ for LaTi$_2$Al$_{20}$.

The strong hybridization effects have been found in this compound such as the $-\ln T$ dependence of $\rho(T)$, a large antiferromagnetic (AF) Weiss temperature $\Theta_{\rm W}$ of 40 K at high $T$s, and the Kondo resonance peak at the Fermi energy \cite{Matsunami}. Moreover,  the sister compound  PrV$_2$Al$_{20}$, whose ground state is also the cubic $\Gamma_3$ doublet, has  an even larger AF $\Theta_{\rm W}= 55$ K and a smaller $T_{\rm Q} = 0.6$ K, and exhibits anomalous metallic behavior suggesting the quadrupolar Kondo effect \cite{akito}. This indicates that the chemical pressure induced by the smaller ionic radius of V than Ti leads to stronger Kondo coupling in PrV$_2$Al$_{20}$ than in PrTi$_2$Al$_{20}$, which suppresses $T_{\rm Q}$, and instead promotes hybridization between the $\Gamma_3$ doublet and conduction electrons. 

While several Pr based superconductors have been reported such as PrOs$_4$Sb$_{12}$ \cite{PrOs4Sb12}, PrRu$_4$Sb$_{12}$ \cite{PrRu4Sb12}, PrPt$_2$B$_2$C \cite{PrPt2B2C}, and PrPt$_4$Ge$_{12}$ \cite{PrPt4Ge12}, all have a CEF ground singlet, except PrIr$_2$Zn$_{20}$ isostructural to PrTi$_2$Al$_{20}$, which has also a nonmagnetic cubic $\Gamma_3$ ground doublet  \cite{Onimaru2,Onimaru3}. In  PrIr$_2$Zn$_{20}$, the superconductivity emerges at 50 mK in the antiferroquadrupolar phase whose ordering temperature is $T_{\rm Q} = 0.11$ K. The critical field has not been precisely determined, but is known to be $< 20$ Oe \cite{PrIr2Zn20.dHvA}. In  contrast with the strong hybridization effects in PrTi$_2$Al$_{20}$, $4f$ electrons are so localized in PrIr$_2$Zn$_{20}$ that do not exhibit the Kondo effect in $\rho(T)$,  show nearly Curie behavior of $\chi(T)$ with a vanishingly small $\Theta_{\rm W}$, and have the cyclotron effective mass of $m_{\rm cyc}^{*}/m_0 \sim 1$ \cite{PrIr2Zn20.dHvA,Onimaru3}. 

Thus, the origin of the higher  $T_{\rm c}$ of the superconductivity in PrTi$_2$Al$_{20}$ than in PrIr$_2$Zn$_{20}$ may partly come from the larger effective mass or density of states of itinerant electrons coming from the stronger hybridization. In this case, the superconducting $T_{\rm c}$ and its associated effective mass would increase under pressure as pressure enhances the hybridization and suppresses the quadrupolar ordering temperature. Therefore, it is highly interesting how the quantum criticality of quadrupolar phase emerges as a function of pressure in the quadrupolar Kondo system PrTi$_2$Al$_{20}$ and its related compounds. 


\begin{acknowledgment}
We thank Y. Karaki, E. C. T. O' Farrell, Y. Uwatoko, K. Matsubayashi, and K. Ueda for useful discussions.
This work was partially supported by Grants-in-Aid (No.21684019) from 
JSPS, by Grants-in-Aids for Scientific Research on Innovative Areas ``Heavy Electrons" of MEXT, Japan and Toray Science Foundation. 
\end{acknowledgment}
\bibliographystyle{jpsj}

\end{document}